\documentclass{ws-procs11x85}
\usepackage{balance} 

\newcommand{\met}{\hbox{E\kern-0.5em\lower-0.1ex\hbox{/}}_T}

\begin{document}

\twocolumn[

\title{OUTFLOWS FROM MAGNETOROTATIONAL SUPERNOVAE}

\author{S.G.Moiseenko$^1$ and G.S.Bisnovatyi-Kogan$^2$}

\address{Space Research Institute,\\
Moscow 117997, Russia\\
$^1$E-mail: moiseenko@iki.rssi.ru\\
$^2$E-mail: gkogan@iki.rssi.ru\\
www.iki.rssi.ru}




\begin{abstract}
We discuss results of 2D simulations of magnetorotational(MR)
mechanism of core collapse supernova explosions. Due to the
nonuniform collapse the collapsed core rotates differentially. In
the presence of initial poloidal magnetic field its toroidal
component appears and grows with time. Increased magnetic pressure
leads to foramtion of compression wave which moves outwards. It
transforms into the fast MHD shock wave (supernova shock wave). The
shape of the MR supernova explosion qualitatively depends on the
configuration of the initial magnetic field. For the dipole-like
initial magnetic field the supernova explosion develops mainly along
rotational axis forming mildly collimated jet. Quadrupole-like
initial magnetic field leads to the explosion developing mainly
along equatorial plane. Magnetorotational instability was found in
our simulations. The supernova explosion energy is growing with
increase of the initial core mass and rotational energy of the core,
and corresponds to the observational data.
\end{abstract}
\vskip12pt  
]

\bodymatter

\section{Introduction}\label{aba:sec1}
The problem of explanation of the core-collapse supernova event is
one of the interesting and long standing problems in astrophysics.
Mechanisms, based on the bounce shock energy or neutrino interaction
with the matter of pre-supernova star do not lead to the supernova
explosion.

The MR mechanism for core collapse supernova explosion was suggested
by Bisnovatyi-Kogan in 1970 \cite{bk1970}.  The main idea of the MR
mechanism is to transform part of the rotational energy of
presupernova into the radial kinetic energy (explosion energy). Due
to the  not uniform collapse the iron core rotates differentially.
Differential rotation leads to the appearing and amplification of
the toroidal component of the magnetic field. Growth of the magnetic
field means amplification of the magnetic pressure with time. A
compression wave appears near the region of the extremum of the
magnetic field. This compression wave moves outwards along steeply
decreasing density profile. In a short time it transforms to the
fast MHD shock wave. When the shock reaches the surface of the
presupernova it ejects part of the matter and energy of the
presupernova star. This ejection can be interpreted as an explosion
of the core collapse supernova star. First 1D simulations of MR
supernova explosion were represented in \cite{bkps}, see also
\cite{bkbook}.

The first   simulations of the MR processes in stars have been done
by \cite{leblanck}, after which MR processes in the stars in
relation to the core collapse supernova explosion  had been
simulated by \cite{bkps, abkp, mueller, ohnishi, symbalisty}.
Recently the interest to the MR processes (especially in application
to the core collapse supernova) was recommenced due to increasing
number of observational data about asymmetry of the explosion, and
possible collimated ejecta in connection with cosmic gamma ray
bursts (\cite{abkm2000, akiyama, yamada, takiwaki, kotake2004,
abkm2005, yamasaki}).

Our results of simulations of MR supernova explosion mechanism show
that this mechanism allows to produce $0.5-1.3\cdot 10^{51}$ergs
energy of explosion. These values of SN explosion energy correspond
to estimations made from core collapse SN observations.

It was found, that the shape of the MR SN explosion qualitatively
depends on the configuration (symmetry type) of magnetic field. The
initial field of quadrupole type of symmetry leads to the MR
explosion which develops predominantly near equatorial plane, while
the initial dipole type field results in SN explosion as mildly
collimated jet developing along axis of rotation.

MR instability was revealed in 2D simulations of MR SN mechanism. MR
instability leads to the exponential growth of both poloidal and
toroidal components of the magnetic field and significantly reduce
time of the MR supernova explosion in comparison with 1D simulations
\cite{bkps}.

\begin{figure}[t]
\center \centerline{\psfig{figure=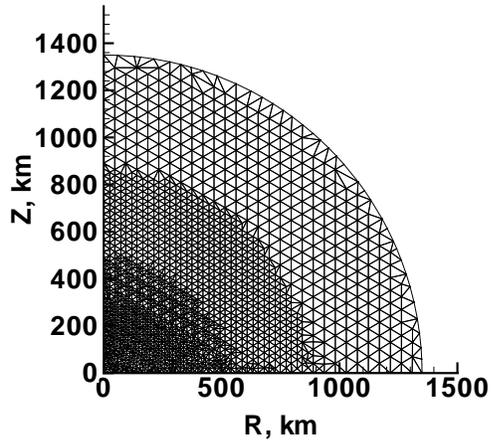,width=7truecm}}
\caption{Example of the triangular Lagrangian grid used for
simulations of MR supernova explosions. \label{grid}}
\end{figure}

\begin{figure}[t]
\center \centerline{\psfig{figure=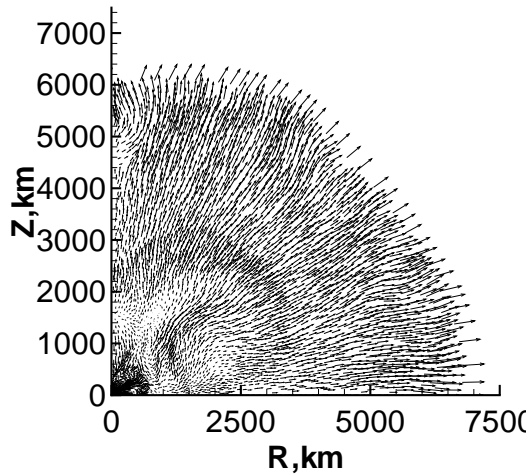,width=7truecm}}
\caption{Example of velocity field for the MR supernova explosion
with the initial quadrupole field. \label{quad_field}}
\end{figure}

\begin{figure}[t]
\center \centerline{\psfig{figure=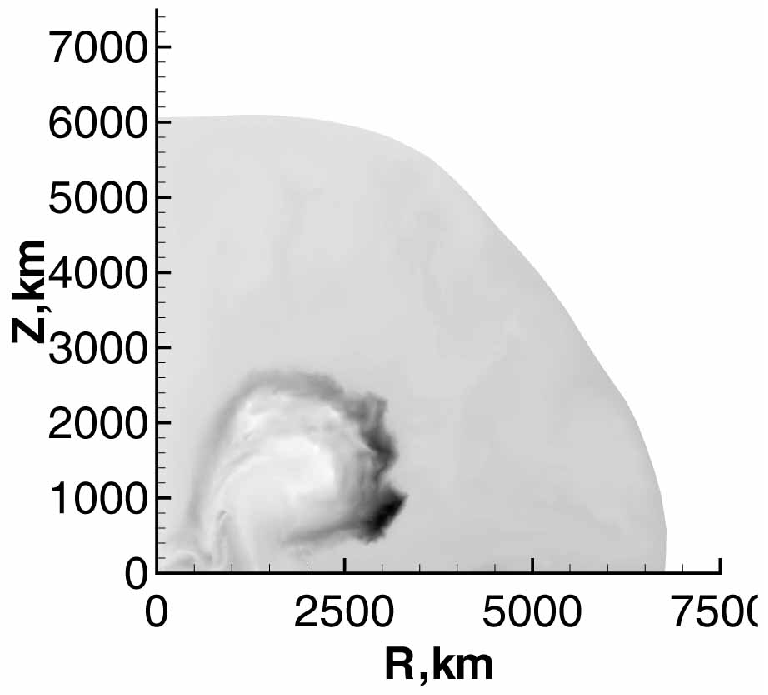,width=7truecm}}
\caption{Example of specific angular momentum distribution for the
MR supernova explosion with the initial quadrupole field.
\label{quad_mom}}
\end{figure}

\begin{figure}[t]
\center \centerline{\psfig{figure=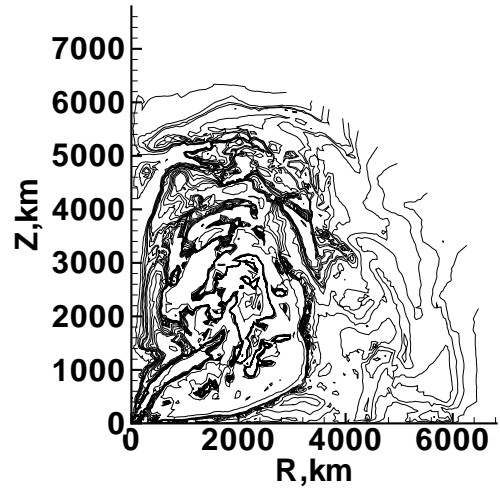,width=7truecm}}
\caption{Example of specific angular momentum distribution for the
MR supernova explosion with the initial dipole field.
\label{dip_mom}}
\end{figure}

For the 2D MHD simulations we used specially developed numerical
code based on the implicit completely conservative Lagrangian scheme
on triangular grid of variable structure (see \cite{ardkos} and
references therein). It  was tested thoroughly with different tests.
The example of the Lagrangean triangular gird is given at the
Fig.\ref{grid}.

Grid reconstruction procedure applied here for the reconstruction of
the triangular lagrangian grid is used both for the correction of
the "quality" of the grid and for the dynamical adaptation of the
grid.

\section{Formulation of the problem}

For the simulations of the magnetorotational supernova with initial
dipole-like magnetic field leading to the formation of protojet, we
used the set of MHD equations with self-gravitation and infinite
conductivity as in the papers \cite{abkm2005, mbka}. We rewrite the
MHD equations in nondimensional form using the same scales as in the
papers \cite{abkm2005, mbka}. We used the same equation of state and
neutrino losses formulae as in the papers \cite{abkm2005, mbka}, and
have started calculations from the same model as in the papers
\cite{abkm2005, mbka}.

The initial poloidal magnetic field is defined as in our previous
paper \cite{abkm2005} by the toroidal current $j_\varphi$ using
Bio-Savara law. The  toroidal current which determines the initial
magnetic field should be defined in the upper and in the lower
hemispheres. The  field with the quadrupole-like symmetry is formed
by the toroidal current antisymmetrical to the equatorial plane. The
dipole-like magnetic field is formed by current symmetrical to the
equatorial plane.
\subsection{Basic equations} Consider a set of
magnetohydrodynamical equations with self-\-gra\-vi\-ta\-tion and
infinite conductivity:
\begin{eqnarray}
\frac{{\rm d} {\bf x}} {{\rm d} t} = {\bf v}, \nonumber \\
\frac{{\rm d} \rho} {{\rm d} t} +
\rho \nabla \cdot {\bf v} = 0,  \nonumber\\
\rho \frac{{\rm d} {\bf v}}{{\rm d} t} =-{\rm grad}
\left(P+\frac{{\bf H} \cdot {\bf H}}{8\pi}\right) + \frac {\nabla
\cdot({\bf H} \otimes {\bf H})}{4\pi} -
\rho  \nabla \Phi, \nonumber\\
\rho \frac{{\rm d}}{{\rm d} t} \left(\frac{{\bf H}}{\rho}\right)
={\bf H} \cdot \nabla {\bf v},\> \Delta \Phi=4 \pi G \rho,
\nonumber\\
\rho \frac{{\rm d} \varepsilon}{{\rm d} t} +P \nabla \cdot {\bf
v}+\rho F(\rho,T)=0,
 \nonumber\\
P=P(\rho,T),\> \varepsilon=\varepsilon(\rho,T). \nonumber
\end{eqnarray}
here $\frac {\rm d} {{\rm d} t} = \frac {\partial} {
\partial t} + {\bf v} \cdot \nabla$ is the total time
derivative, ${\bf x} = (r,\varphi , z)$, ${\bf
v}=(v_r,v_\varphi,v_z)$ is the velocity vector, $\rho$ is the
density, $P$ is the pressure,  ${\bf H}=(H_r,\> H_\varphi,\> H_z)$
is the magnetic field vector, $\Phi$ is the gravitational potential,
$\varepsilon$ is the internal energy, $G$ is gravitational constant,
${\bf H} \otimes {\bf H}$ is the tensor of rank 2, and $F(\rho,T)$
is the rate of neutrino losses.

$r$, $\varphi$, and  $z$ are spatial Lagrangian coordinates, i.e.
$r=r(r_0,\varphi_0,$ and $z_0,t)$,
$\varphi=\varphi(r_0,\varphi_0,z_0,t)$, and
$z=z(r_0,\varphi_0,z_0,t)$, where $r_0,\varphi_0,z_0$ are the
initial coordinates of material points of the matter.

Taking into account symmetry assumptions ($ \frac \partial {\partial
\varphi} = 0$), the divergency of the tensor ${\bf H} \otimes {\bf
H}$ can be presented in the following form:
$$
{\rm \nabla\cdot}({\bf H} \otimes {\bf H})= \left(\begin{array}{l}
\frac {1}{r} \frac {\partial(rH_rH_r)}{\partial r} + \frac
{\partial(H_zH_r)} {\partial z}-
\frac {1}{r} H_\varphi H_\varphi \\
\frac {1}{r} \frac {\partial(rH_rH_\varphi)}{\partial r} + \frac
{\partial(H_zH_\varphi)} {\partial z}+
\frac {1}{r} H_\varphi H_r \\
\frac {1}{r} \frac {\partial(rH_rH_z)}{\partial r} + \frac
{\partial(H_zH_z)} {\partial z}
\end{array}
\right).
$$

Axial symmetry ($\frac \partial {\partial \varphi}=0$) and symmetry
to the equatorial plane  are assumed. The problem is solved in the
restricted domain. At $t=0$ the domain is restricted by the
rotational axis $r\geq 0$, equatorial plane $z\geq 0$, and the outer
boundary of the star where
 the density of the matter is zero, while poloidal
components of the magnetic field $H_r$, and $H_z$ can be non-zero.

 At
the rotational axis ($r=0$) the following boundary conditions are
defined: $(\nabla \Phi)_r=0,\> v_r=0$. At the equatorial plane
($z=0$) the boundary conditions are: $(\nabla \Phi)_z=0,\> v_z=0$.
At the outer boundary (boundary with vacuum) the following condition
is defined:
 $P_{\textrm {outer boundary}}=0$.

\section{MR supernova explosion}

MR explosion with the initial quadrupole-like magnetic field was
described in detail in the paper \cite{abkm2005}. After the core
collapse the pre SN rotates differentially. The toroidal component
of the magnetic field appears and grows linearly with time at the
initial stage of the MR explosion. When the toroidal magnetic field
reaches some certain  value its linear growth changes to the
exponential growth of the toroidal and poloidal components due to
the appearing of the magnetorotational instability (MRI).  A toy
model for the qualitative explanation of the MRI in MR supernova was
suggested in \cite{abkm2005}. MR supernova explosion with initial
quadrupole-like magnetic field results in explosion which develops
mainly near equatorial plane. At the Fig.\ref{quad_field} the
velocity field  and specific angular momentum Fig.\ref{quad_mom} for
the initial quadrupole-like field.

Simulations of the MR supernova with the initial magnetic field of
dipole-like symmetry leads to the qualitatively different result in
the shape of explosion \cite{mbka}. In this case the MR explosion
develops mainly along the axis of rotation and forms mildly
collimated proto jet (Figure \ref{dip_mom}).
The protojet found in our simulations
could be collimated when it develops in the extended envelope of the
massive star - the progenitor of the core collapse supernova.

MRI leads to the formation of chaotic magnetic field structure. I
the case of finite conductivity the reconnection of the magnetic
field could be important. We have estimated characteristic time of
the reconnection of the magnetic field using results of our
simulations \cite{mbka}. We found that characteristic time for the
reconnection of the magnetic field for supernova parameters used in
our simulations is $\approx 5$s. The MR supernova explosion time in
our simulations is $\approx 0.5-1$s what is significantly less then
characteristic time of the magnetic field reconnection development,
and do not influence the MR supernova significantly.

\begin{figure}[t]
\center \centerline{\psfig{figure=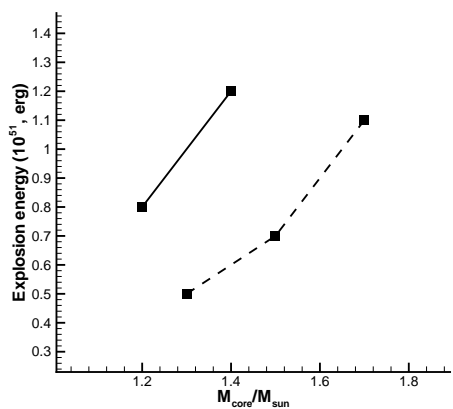,width=7truecm}}
\caption{Dependence of the supernova explosion energy on the core
mass for initial angular velocity $\omega_0$ $\approx
3.53s^{-1}$(solid line) and $\omega_0$ $\approx 2.52s^{-1}$(dashed
line) (before collapse) \label{expl_en}}
\end{figure}

We have done simulations for  different initial masses of the iron
core and for  different initial angular velocities of presupernova.
The supernova explosion energy grows significantly with increase of
the core mass. The dependence of the explosion energy on the core
mass for the different initial values of the rotational energy
(angular velocity) is presented at Fig.\ref{expl_en} for the initial
quadrupole-like magnetic field.

\section{Conclusions}
The results of 2D simulations of MR supernova explosion mechanism
show that it allows to get the explosion energy which corresponds to
observational values of the explosion energy for core collapse
supernovae. The magnetorotational instability appearing in the
simulations of MR supernova significantly reduces the explosion
time, and increase the chaotic magnetic field in the young neutron
stars \cite{abkm2005}. We have found that MR supernova explosion
energy is growing with increase of the initial mass of the iron core
and initial rotational energy.

This work was partially supported by RFBR grants 05-02-17697A,
06-02-91157 and 06-02-90864 and President's grant for leading
scientific schools No.NS 10181.2006.2


\end{document}